
\magnification 1200
\baselineskip=0.5truecm
\hsize=15.4truecm
\vsize=24truecm
\topskip=1truecm
\hfill$DIAS-STP-94-19$
\raggedbottom
\abovedisplayskip=3mm
\belowdisplayskip=3mm
\abovedisplayshortskip=0mm
\belowdisplayshortskip=2mm
\normalbaselineskip=12pt
\font\grand=cmbx10 at 14.4truept
 at 12.0truept
\def\dslash{D \kern-6pt /}

\def\p1{p_{{1 \over 2}}}
\def\p3{p_{{3 \over }}}
\def \lwd#1{\lower2pt\hbox{$\scriptstyle #1$}}

\overfullrule=0pt
\rightline{ }
\vskip 0.5truecm
\centerline {\grand Non-Trivial Non-Canonical W-Algebras}
\centerline {\grand from Kac-Moody Reductions.}
\vskip 0.7truecm
\centerline {G.A.T.F. da Costa and L. O'Raifeartaigh}
\centerline {Dublin Institute for Advanced Studies}
\centerline {10, Burlington Road, Dublin 4}
\vskip 0.7truecm
{\bf Abstract} {\it By reducing a split $G_2$ Kac-Moody
algebra by a non-maximal set of first-class constraints we produce
W-algebras which (i) contain fields of negative
conformal spin and (ii) are not trivial extensions of canonical
W-algebras.}

\vskip 0.3truecm
{\bf Introduction.} In recent years it has been found that
W-algebras occur naturally in the reduction of Poisson-bracket
Kac-Moody (KM) algebras
by first-class constraints [1] [2], where the constraints consist of
reducing a nilpotent subalgebra $\Gamma$ of the KM algebra to a single
constant nilpotent generator $M_-$. With each such nilpotent element
$M_-$ is associated an $sl(2)$ subalgebra ${\cal S}$ of the
underlying Lie algebra, and to date almost all of the W-algebras obtained
by such reductions have been characterized by the fact that they
have a basis corresponding to the highest weights of ${\cal S}$.
Such W-algebras will be called {\it canonical} and
they correspond to constraint algebras $\Gamma_c$ which are
(i) positively graded with respect to ${\cal S}$ and (ii)
have maximal dimension subject to the conditions  that
the constraints be first-class (see (2) below).

The remainder of the W-algebras that have been obtained by
KM reduction are direct sums of canonical W-algebras and
free-field algebras. For lack of a better name we shall call these
W-algebras {\it quasi-canonical}. The corresponding constraint
subalgebras $\Gamma_q$ are subalgebras of the canonical constraint
subalgebras $\Gamma_c$ in which some or all of the elements in the lowest
grade are ommitted. Examples are the constraint algebras
[3] in which all of the grade ${1\over 2}$ elements of
$\Gamma_c$ are omitted.

  The choices of $\Gamma_c$ and $\Gamma_q$ as constraint
algebras $\Gamma$ were made in the previous reductions
because they guarantee that the reduced
KM algebra will be a W-algebra i.e. will be differential polynomial
and will have a basis consisting of a Virasoro and primary fields.
However, although these choices are {\it sufficient} to guarantee
this it is not clear that they are
{\it necessary}. In ref. [4] the necessary conditions
were investigated and some strong lower bounds on the dimensions
of potential $\Gamma$ subalgebras were found.
These lower bounds fall short of requiring that $\Gamma\supseteq
\Gamma_q$ and thus create
a margin for construction of a new kind of W-algebra by reduction.
We do not believe that the margin is very large but we wish to show in
this paper that it is at any rate not empty, by constructing two such
W-algebras. The two W-algebras are obtained by the reduction of a split
$G_2$ KM algebra with a $(3+2 \times 4+3\times 1)$ embedding, using
constraint algebras of the form
$\hat \Gamma_c=\Gamma_c/\bar \Gamma_{{3 \over 2}}$ and
$\hat \Gamma_q=\Gamma_q/\bar \Gamma_{{3 \over 2}}$,
where the quotient is with respect to a 1-parameter invariant
subalgebra of grade ${3 \over 2}$. The W-algebras obtained in this way differ
from the canonical
ones in that \noindent (i) they contain a primary field of strictly
negative (minus one-half) conformal spin
\noindent (ii) the subalgebras $W_d$ with the
spin-content of the canonical algebra $W_c$ are non-trivial
deformations of $W_c$ and
(iii) $W_d$ does not decouple from its complement and in particular
does not decouple from the negative spin field. The existence
of a coupled field with negative  conformal spin is rather unexpected and
raises
some questions about the unitarity of the quantized version [4].

{\bf The $G_2$ KM algebra and Constraints: } We use
the conventional root diagram for $G_2$ and consider the
horizontal $sl(2)$-embedding i.e. the embedding with irreps
of dimension $\{1,4,(3,1),4,1\}$. We denote the
 generators of this $sl(2)$ by $\{M_-,M_0,M_+\}$
with Lie algebra $[M_0,M_{\pm}]=\pm M_{\pm}$,$[M_-,M_+]=M_0$.
For $G_2$ there is an orthogonal (vertical) $sl(2)$ with generators
$\{Y_-,Y_0,Y_+\}$ and the same Lie algebra.
The remaining roots (which form two quadruplets with respect to the
$M$'s and four doublets with respect to the $Y$'s) can then be
labelled $E^y_m$ where $m$ and $y$ are the eigenvalues of $M_0$
and normalized so that $[M_+,E^y_m]=-E^y_{m+1}$ and $[M_-,E^y_m]={1
\over 2}(j+m)(j-m+1)E^y_{m-1}$ for $j={3 \over 2}$ and similarly for
the $Y$'s with $j={1 \over 2}$. A normalization
of the roots compatible with these conventions is
$$<M_+,M_->=1 \qquad <M_0,M_0>=-1   \qquad
<E^y_{-m},E^{-y}_m>=(-1)^{m-{1 \over 2}}   \eqno(1)$$
with the $Y$'s normalized in the same way as the $M$'s.
\vskip 0.2truecm
 We denote the $G_2$ KM fields corresponding to the horizontal
and vertical $sl(2)$'s by $\{j_-(x),j_0(x),j_+(x)\}$ and $\{\bar
s(x),y(x),s(x)\}$ respectively and the KM fields
corresponding to the roots $E^{\pm {1 \over 2}}_m$
by $t_m(x)$ and $\bar t_m(x)$ respectively. Thus the KM
current $j(x)$ can be written as
$$j_-(x)M_-  +j_0(x)M_0+j_+(x)M_++\bar
s(x)Y_-+y(x)Y_0+s(x)Y_+ +t_m(x)E^{{1 \over 2}}_m+\bar t_m(x)E^{-{1
\over 2}}_m  \eqno(2)$$
which we can represent diagrammatically as
$$\eqalign{
s(&x)  \cr
t_{-{3 \over 2}}(x) \qquad t_{-{1 \over 2}}(x) \quad &
\quad t_{{1 \over 2}}(x) \qquad t_{{3 \over 2}}(x)    \cr
j_-(x) \qquad  j_0(x)&y(x) \qquad  j_+(x)  \cr
\bar t_{-{3 \over 2}}(x) \qquad \bar t_{-{1 \over 2}}(x) \quad &
\quad \bar t_{{1 \over 2}}(x) \qquad \bar t_{{3 \over 2}}(x)    \cr
\bar s(&x)    \cr   }    \eqno(3)$$
\vskip 0.3truecm
 For the above KM-algebra the constraints mentioned in the
Abstract are of the form
$$j_-(x)\equiv <M_+,j(x)>=1   \qquad  <\tilde \gamma,j(x)>=0 ,
\eqno(4)$$
where the constraint algebra $\Gamma$ is a semi-direct sum of the
form $\Gamma=M_+ \wedge \tilde \Gamma$ and $\tilde \gamma$ is any
element of $\tilde \Gamma$. In order that the constraints (1) be
first-class the constraint algebra $\Gamma$ must satisfy [5] the
following conditions
$$<\Gamma,\Gamma>=0 \qquad w(\Gamma,\Gamma)\equiv
<M_-,[\Gamma,\Gamma]>=0 . \eqno(5)$$
The non-trivial components of the current then lie in
$\Gamma^{\perp}$ which, on account of the second condition in (2)
contains $[M_-,\Gamma]$. It is assumed that $M_-$ is
non-degenerate on $\Gamma$ so that $[M_-,\Gamma]$ has the same
dimension as $\Gamma$, and then the natural gauge-fixing procedure
(called the Drinfeld and Sokolov or DS procedure) is to set the components
of the current which lie in $[M_-,\Gamma]$ equal to zero. In other words
the DS gauge-fixing consists in supplementing the constraints (2)
with the further linear constraints
$$<\theta_a,j(x)>=0  \qquad \hbox{where} \qquad w(\theta_a,\gamma_b)
\equiv <\theta_a,[M_-\gamma_b]>=\delta_{ab}
.  \eqno(6)$$
The conformal invariance of the reduction (4) is established by
noting that, although the first constraint in (4) is not compatible
with the conformal group generated by
the usual KM Sugawara Virasoro-operator $L_{KM}(j(x))$, it is
compatible with the conformal group generated by the modified
 Virasoro operator
$$\Lambda(j(x))=L_{KM}(j(x))-h'(x)  \qquad \hbox{where} \qquad
h(x)=-j_0(x) ,  \eqno(7)$$
the sign of $h(x)$ being chosen for later convenience.
With respect to the modified Virasoro operator (7) all of the KM fields
are primary (with conformal weight $(m+1)$) except $h(x)$, whose
conformal variation is of the form
$$\delta_c h(x)=f(x)h'(x)+f'(x)h(x)-kf''(x) ,   \eqno(8)$$
where $f(x)$ is the conformal parameter and $k$ is a constant
proportional to the KM centre, which for convenience we normalize to
$-1$. Thus $h(x)$ transforms as the component of a spin-one connection,
the inhomogeneous part of the conformal variation being
$\delta_I h(x)=f''(x)$.
\vskip 0.3truecm
The different reductions for the chosen $sl(2)$ embedding
are characterized by the different choices of the constraint
subalgebra $\Gamma$. In order to guarantee (5) and to obtain a differential
polynomial W-algebra it is usual to choose $\Gamma$ positive with respect
to $M_0$ and for simplicity we shall follow this procedure. We shall
also assume for simplicity that $\Gamma$ has an $sl(2)$ basis i.e.
has a basis labelled by the $sl(2)$ Casimir as well as $M_0$.
There are then five possibilities, which can be grouped into three
classes as follows:
\vskip 0.1truecm
$$\eqalign{
&\hbox{Class I}  \hskip 1 truecm  \Gamma_c=\{ M_+,\gamma_{{3 \over 2}},
\gamma_{{1 \over 2}},
\bar \gamma_{{3 \over 2}} \}
 \quad \hbox{and} \quad \Gamma_q =
\{ M_+,\gamma_{{3 \over 2}},\bar \gamma_{{3 \over 2}} \}    \cr
&\hbox{Class II} \hskip 0.9truecm  \hat \Gamma_c =
\{ M_+,\gamma_{{3 \over 2}}, \gamma_{{1 \over 2}} \}
\hskip 1truecm \hbox{and} \quad \hat
 \Gamma_q=\{ M_+,\gamma_{{3 \over 2}} \}   \cr
&\hbox{Class III} \hskip 0.8truecm   \Gamma \hskip 0.1truecm =\hskip
0.2truecm M_+  \hskip 1truecm .    \cr
  }  \eqno(9)$$
where $\gamma_m=E^{{1 \over 2}}_m$ for $m={1 \over 2}$, ${3 \over
2}$ and $\bar \gamma_{{3 \over 2}}=E^{-{1 \over 2}}_{{3 \over 2}}$.
Note that the Class II subalgebras are obtained from the respective
Class I subalgebras by omitting the
base-element $\bar \gamma_{{3 \over 2}}$.
\vskip 0.1truecm
 Classes I and III have already been discussed in detail in the
literature. Class I contains the canonical and
quasi-canonical W-algebras for this particular $sl(2)$ embedding.
The canonical W-algebra $W_c$ is generated by $\{W_1,W_2,W_{{5 \over 2}}\}$
where $W_1$ is an $sl(2)$ KM subalgebra generated by the spin-1
fields $\{\bar
s,y,s\}$, $W_2$ is the Virasoro
and $W_{{5 \over 2}}$ is a spin ${5 \over 2}$ doublet.
The quasi-canonical subalgebra is generated by $\{W_{{1 \over 2}},
W_c\}$, where $W_{{1\over 2}}$ is the spin ${1
\over 2}$ doublet generated by the free-fields
$\{t_{-{1 \over 2}},\bar t_{-{1 \over 2}}\}$. The model of Class III is
distinguished by the fact that it does {\it not} produce a W-algebra (for
reasons that will be explained below).
 The question is: what happens for the
models of the intermediate class II? For these models the first-class
constraints are
$$j_-(x)=1 \quad \bar t_{-{3 \over 2}}(x)=\bar t_{-{1 \over 2}}(x)=0
\qquad \hbox{and} \quad j_-(x)=1 \quad \bar t_{-{3 \over 2}}(x)=0
\eqno(10)$$
respectively.
 What we shall show is that both these models produce
W-algebras. But they differ from the W-algebras obtained previously
because they each contain the field $t_{-{3 \over 2}}(x)$, which has
conformal weight minus one-half, and
because the W-algebra is not a direct sum of the canonical W-algebra
and a complementary subalgebra (much less a free-field algebra).
Indeed it does not even contain the canonical W-algebra.  Accordingly,
the models of Class II furnish examples of $W$-algebras which have
negative spins and are not extensions of the canonical $W$-subalgebras.
\vskip 0.3truecm
{\bf Gauge-Fixed Fields:} We first carry out the DS
gauge-fixing for the models of Class II i.e. we gauge-transform to zero
the part of the current that lies in $[M_-,\Gamma]$.
As usual we begin by gauge-transforming  the coefficient $h(x)$ of
$M_0$ to zero by means of the gauge-transformation $e^{h(x)M_+}$.
Under this gauge-transformation the fields change as follows:
$$\eqalign{ &s(x)
\quad y(x) \quad \bar s(x) \quad t_{-{3 \over 2}}(x) \quad \bar
t_{-{3 \over 2}}(x)  \quad \hbox{remain unchanged} \cr &
j_+(x)\rightarrow  w(x)\equiv j_+(x) +h'(x)-{h^2(x) \over 2}  \cr &
t_m(x)  \rightarrow  u_m(x)\equiv
\sum_{p\geq 0}{h^p \over p!}t_{m-p} \quad m\not={-3 \over 2},  \cr  }
\eqno(11)$$
and similarly for $\bar t_m(x)$. From the conformal variation of $h(x)$
given above it is easy to see that the non-primary part of the
conformal variation of these fields are
$$\delta w(x)=f'''(x) \qquad
\delta u_m(x)=f''(x)u_{m-1}(x) , \eqno(12)$$
and similarly for the
$\bar u_m(x)$. Thus $w(x)$ transforms like a spin-2 connection while
the $u$'s and $\bar u$'s transform homogeneously but not primarily.
\vskip 0.2truecm
 We then make gauge-transformations with respect to
$\gamma_{{1 \over 2}}$ and $\gamma_{{3 \over 2}}$. In preparation
for this we summarize the relevant $G_2$ commutation relations, namely
$$\eqalign{
[\gamma_{{1 \over 2}},M_-]=-2E^{{1 \over 2}}_{-{1 \over 2}} \qquad
[\gamma_{{1 \over 2}},E^{{1 \over 2}}_{-{1 \over 2}}]&=-{1 \over
2}Y_+ \qquad
[\gamma_{{1 \over 2}}[\gamma_{{1 \over 2}},M_-]]=Y_+   \cr
[\gamma_{{1 \over 2}},Y_0]=-{1 \over 2}\gamma_{{1 \over 2}} \qquad &
\qquad [\gamma_{{1 \over 2}},Y_-]=-{1 \over 2}E^{-{1 \over 2}}_{{1
\over 2}}   \cr  }      \eqno(13)$$
and
$$[\gamma_{{3 \over 2}},M_-]=-{3 \over 2}\gamma_{{1 \over 2}} \qquad
[\gamma_{{3 \over 2}},E_{-{3 \over 2}}^{{1 \over 2}}]={1 \over
2}Y_+  \qquad [\gamma_{{3 \over 2}},Y_-]=
-{1 \over 2}E^{-{1 \over 2}}_{{3 \over 2}}  \eqno(14)$$
These and all other $G_2$ commutation relations can be
obtained by using the normalizations of the $G_2$ generators given
in the first paragraph and the complete anti-symmetry of $<X[Y,Z]>$.

 From the above relations we see that with respect to
gauge-transformations generated by $\alpha\gamma_{{1 \over 2}}$ and
$\beta\gamma_{{3 \over 2}}$ we have
$$u_{-{1 \over 2}} \quad \rightarrow \quad u_{-{1 \over 2}}-2\alpha
\hskip 2truecm  u_{{1 \over2}}\quad \rightarrow \quad
u_{{1 \over 2}}-(\partial +{y \over 2})\alpha -{3 \over 2}\beta
\eqno(15)$$
Since the DS gauge fixing in model $\hat \Gamma_c$ consists of
gauging $u_{-{1 \over 2}}$ and $u_{{1 \over 2}}$ to zero we see that
the appropriate gauge-parameters are
$$\alpha^c={1 \over 2}u_{-{1 \over 2}} \qquad \hbox{and} \qquad
\beta^c={2 \over 3}v^c_{{1 \over 2}} \qquad \hbox{where} \qquad v^c_{{1 \over
2}}=u_{{1 \over 2}}-{1 \over
2}(\partial + {y \over 2})u_{-{1 \over 2}}  \eqno(16)$$
Making a gauge-transformation with these parameters we find that the
fields in the DS gauge for this model are
$$\eqalign{
&s^c  \cr
t_{{3 \over 2}} \qquad 0 \quad & \quad  0 \qquad v^c_{{3 \over 2}}
\hskip 4.7truecm s^c=s-{1 \over 4}u^2_{-{1 \over 2}}+{1 \over 3}
t_{-{3 \over 2}}v^c_{{1 \over 2}}    \cr
1 \qquad \quad  & y \qquad w^c
\hskip 2.5 truecm  \hbox{where} \hskip 1.5truecm \bar v^c_{{1 \over 2}}=\bar
u_{{1 \over
2}}-{1 \over 4}u_{-{1 \over 2}}\bar s      \cr
0 \qquad 0 \quad & \quad \bar v^c_{{1 \over 2}} \qquad \bar v^c_{{3 \over 2}}
\hskip 4.5truecm \bar v^c_{{3 \over 2}}=\bar u_{{3 \over 2}}-{1 \over
3}\bar s v^c_{{1 \over 2}}     \cr
&\bar s   \cr   }   \eqno(17)$$
It is straightforward to compute $w^c$ and $v^c_{{3 \over 2}}$
also but, except for the part that will be
computed independently below their precise forms  will not be needed.

For the model $\hat \Gamma_q$ the gauge-parameter
$\alpha$ is identically zero and the DS gauge-fixing consists only
of gauging $u_{{1 \over 2}}$ to zero. It is clear that the
appropriate value of the gauge-parameter $\beta$ in this case is
$\beta^c={2 \over 3}u_{{1 \over 2}}$ and after the
gauge-transformation with this value of $\beta$ we obtain the DS
fields
$$\eqalign{
&s^q  \cr
t_{{3 \over 2}} \qquad u_{-{1 \over 2}} \quad & \quad 0 \qquad v^q_{{3 \over
2}}
\hskip 4.7truecm
s^q=s+{1 \over 3}t_{-{3 \over 2}}u_{{1 \over 2}}                    \cr
1 \qquad \quad & y \qquad w^q  \hskip 2.5truecm \hbox{where}    \cr
0 \qquad \bar t_{-{1 \over 2}} \quad & \quad \bar u_{{1 \over 2}}
\qquad \bar v^q_{{3 \over 2}}   \hskip 4.5truecm  \bar v^q_{{3 \over 2}}=\bar
u_{{3 \over 2}}-{1 \over 3}\bar s u_{{1 \over 2}}      \cr
&\bar s   \cr   }   \eqno(18)$$
Again it is straightforward to compute $w^q$ and $v^q_{{3 \over 2}}$
but their precise forms will not be needed.
In each of the above models the final fields provide a complete
basis for the gauge-invariant functions of the KM fields
and hence the Poisson-bracket algebra of these fields induced by
the original KM algebra closes. We call this algebra the reduced
algebra. Since the fields are differential polynomials
of the original KM fields it is obviously a differential polynomial
algebra. Thus it will be a W-algebra if it has a basis consisting of a
Virasoro operator and a set of fields which are primary with respect
to it. For short we shall call such a basis a primary basis.
 \vskip 0.3truecm
 {\bf Existence of a Primary Basis:} We wish to show that
for the two models of Class II the reduced
algebra has a primary basis. First we note that the modified Virasoro operator
$\Lambda(x)$ of (4) is a candidate for the Virasoro base-element
since it is a differential polynomial in the KM current components
and  transforms in the correct manner. Furthermore, since
$\Lambda(x)$ and $W(x)$ are the only
gauge-fixed fields which are linear in $j_+(x)$ either one can be
chosen as a base element of the reduced algebra
(which implies, of course, that $\Lambda(x)$ and $W(x)$ differ only
by a differential polynomial in the other gauge-fixed  fields and
it is easy to verify that such is indeed the case). Because of its
transformation properties we choose $\Lambda(x)$.
\vskip 0.2truecm
 To obtain some orientation for considering the primariness
of the other base elements in
the models of Class II we first consider the single model of Class
III, which contains all the $u$-fields. The $u$'s are the only fields
which are not primary and even for them the fields $u_{-{1 \over 2}}$,
$u_{{3 \over 2}}$ and the corresponding barred fields can be converted
to primary fields by the addition of differential polynomials in
fields of lower grades, namely
$$u_{-{1 \over 2}} \rightarrow p_{-{1 \over 2}} \equiv
u_{-{1 \over 2}}+2\partial t_{-{3 \over 2}}   \eqno(19)$$
$$u_{{3 \over 2}}\rightarrow p_{{3 \over 2}}\equiv
u_{{3 \over 2}}-{2 \over 3}\partial u_{{1 \over 2}}
-{1 \over 6}(2\partial^2+3\Lambda)u_{-{1 \over 2}} +{1 \over 9}
\bigl(2\partial^3+7\Lambda\partial + 4(\partial \Lambda)\bigr)t_{-{3 \over 2}}
,
    \eqno(20)$$
and similarly for the barred fields. The non-primariness of the
model comes from the $u_{{1 \over 2}}$ field and its barred
counterpart. The point is that, so long as $t_{-{3 \over 2}}$
is not zero, $u_{{1 \over 2}}$ can not be converted to a primary field.
To see this we note from (12) that the $u$-fields
and $w$ transform {\it linearly} with respect to the conformal group
and $w$ transforms linearly with respect to the Mobius subgroup
($f'''(x)=0$). Hence $u_{{1 \over 2}}$ must be converted to a primary field
even at the linear level. This can be done only
by the addition of terms that are {\it linear} in the $u$'s and $w$
and since the only fields with the correct quantum
numbers are $u'_{-{1 \over 2}}$ and $t''_{-{3 \over 2}}$, of
which $t''_{-{3 \over 2}}$ is Mobius-primary,  the only possibility
is to let $u_{{1 \over 2}}\rightarrow
U_{{1 \over 2}}=u_{{1 \over 2}}+\kappa u'_{-{1 \over 2}}$
where $\kappa$ is a constant. But from (12) it follows
that the non-conformal part of the variation of $U_{{1 \over 2}}$
with respect to the Mobius subgroup is
$\delta U_{{1 \over 2}}= \Bigl((1 +{\kappa \over 2})u_{-{1
\over 2}}+\kappa t'_{-{3 \over 2}}\Bigr)f''$, which shows that no choice of
$\kappa$ can make $U_{{1 \over 2}}$ primary and thus convert $u_{{1
\over 2}}$ to a primary field.
Similarly, unless $\bar t_{-{3 \over 2}}$ is zero,  the field $\bar
u_{{1 \over 2}}$ cannot be converted to a Mobius-primary field, much
less a totally primary one.
 This is why the reduced algebra of model III, for which $u_{{1
\over 2}}$ and $\bar u_{{1 \over 2}}$ are base elements, is not a
$W$-algebra. (The argument that primariness must hold at the linear
level alone is given in more detail in [3] and the above result
is actually a special case of the result that $u_n$ cannot be made
even Mobius-primary unless $u_{-(n+1)}=0$).
\vskip 0.3truecm
 In both models of Class II the field $\bar t_{-{3 \over 2}}$ is zero
so there is no problem in converting the field $\bar u_{{1 \over 2}}$
 to a primary field, namely by letting
$$\bar u_{{1 \over 2}}\rightarrow \bar p_{{1 \over 2}}\equiv
u_{{1 \over 2}}-2\partial \bar t_{-{1 \over 2}} .  \eqno(21)$$
The field $u_{{1 \over 2}}$ remains non-primary but it is no longer  a
 base-element of the reduced algebra because it is gauged away
 by the DS gauge-fixing corrresponding to
$\gamma_{{3 \over 2}}$. Of course, it reappears in other elements of
the reduced algebra, but, as can be seen by inspection of (15)
through (18) it appears in these elements only in the combinations
$$u_{{3 \over 2}}-{2 \over 3}u'_{{1 \over 2}} \quad (\hbox{in}
\hskip 0.2truecm  v_{{3 \over 2}})
\qquad u_{{1 \over 2}}t_{-{3 \over 2}} \quad (\hbox{in}\hskip 0.2truecm s)
\qquad u_{{1 \over 2}} y \quad  (\hbox{in}
\hskip 0.2truecm  v_{{3 \over 2}})
\qquad
 u_{{1 \over 2}}\bar s \quad (\hbox{in}\hskip 0.2truecm   \bar v_{{3 \over 2}})
   \eqno(22)$$
We now wish to show that, because $u_{{1 \over 2}}$  appears only in these
combinations, the DS basis can be replaced by a basis of primary fields.
For the first expression in (22) the conversion is
already given by (20). For the second expression the conversion is easily
verified to be
$$u_{{1 \over 2}}t_{-{3 \over 2}}\rightarrow p_0\equiv u_{{1 \over 2}}t_{-{3
\over 2}}+2u_{-{1 \over 2}}\partial t_{-{3 \over 2}}+2(\partial
t_{-{3 \over 2}})^2 .    \eqno(23)$$
The last two expressions in (22) can be handled together by using the
general observation that if
$T_m$ is any primary field of grade $m$ (spin $m+1$), such that
$m\not=-1$ or $-{3 \over 2}$ then the modified field $P_m$  given by
$$P_m\equiv u_{{1 \over 2}}T_m -{1 \over m+1}u_{-{1 \over 2}}\partial
T_m +{1 \over (m+1)(2m+3)}t_{-{3 \over
2}}(\partial^2-(m+1)\Lambda)T_m ,   \eqno(24)$$
is primary. The formula for $m=0$ obviously applies to the last
two expressions in (21) and thus converts these two expressions into primary
fields. Since in all the above conversions only gauge-invariant
fields were used and the highest spin field in each case remained
unchanged it is clear that the resulting primary fields (together with the
modified
Virasoro) form a basis. We have thus established by construction
that for both models of Class II there is a primary basis. The
construction is rather ad hoc and it would be interesting to find a
more systematic way of obtaining the primary basis.
\vskip 0.5truecm
{\bf General Structure of the W-Algebras.}  Having established
that there exists a primary  basis for the DS gauge-fixed fields we now
wish to investigate the Poisson-bracket (PB) algebra of these fields. We
first note that the DS fields are gauge-invariant polynomials of the
original KM fields which reduce to independent KM fields in the
linear limit. This means that they form a basis for all the
gauge-invariant differential polynomials of the KM fields and that
the reduced algebra is a freely-generated differential polynomial
algebra. By its PB-structure we shall mean the PB structure
of the differential polynomials inherited from the KM PB structure
$$\{<A,j(x)>,<B,j(y)>\}=<[A,B],j(x)>\delta(x-y) +c<A,B>\delta'(x-y)
\eqno(25)$$
The spin structures of the DS basis for the models
$\hat \Gamma_c$ and $\hat \Gamma_q$ are  of the form
$$W=W_d+(W_{-{1 \over 2}} +W_{{3 \over 2}})  \quad \hbox{and} \quad
W=W_d+W_{{1 \over 2}}+(W_{-{1 \over 2}}+W_{{3 \over 2}})
\eqno(26)$$
respectively, where $W_d$ is a 6-dimensional subalgebra with the spin
content of the canonical W-algebra and $W_{{1 \over 2}}$ is
a 2-dimensional subalgebra with the spin content of the extra fields
that appear in the quasi-canonical model. $W_{-{1 \over 2}}$ and
$W_{{3 \over 2}}$ are 1-dimensional and contain the two extra fields
that appear in the present models.  As we shall see later in (37) the
pair of fields
$\{u_{-{1 \over 2}},\bar t_{-{1 \over 2}} \}$ in $W_{{1 \over 2}}$ and
the pair of fields $\{t_{-{3 \over 2}},\bar v_{{1 \over 2}}\}$ which appear
in the present models are each a conjugate pair
with respect to the PB's. Thus the $W$-algebras consist of $W_d$ plus
one and two conjugate pairs respectively. It would be
straightforward to compute the W-algebra in the DS basis but
since the DS basis does not exhibit the direct sum structure in the
quasi-canonical case it probably does not exhibit the true structure
in the present case.
On the other hand, the large freedom in choosing a general basis
(as exemplified by (28) and (29) below) makes it difficult
to extract any detailed information. Hence we restrict
ourselves to obtaining two key structural results, namely that
$W_d$ is a non-trivial deformation of a canonical subalgebra
and that it does not decouple from its complement.

 In order to establish these results we note that since
they are negative it is sufficient to establish them for any subset
of field configurations. Hence, for simplicity, we
restrict ourselves to the {\it constant} configurations.
This has the advantage that all computations become Lie-algebraic,
modulo overall delta-functions which we suppress from now on.

 We also recall from [4] that in the case of the
quasi-canonical subalgebra the free field algebra $W_{{1 \over 2}}$
does not decouple from $W_c$ in the original DS basis but
only in a modified basis. For example, the modification required
for $W_{{1 \over 2}}$ to decouple from the $W_1$ part of the
quasi-canonical model is to change the basis of $W_1$ from
$\{\bar s,y,s\}$ to $\{\bar s,y,s\}-{1 \over 4}\{\bar t_{-{1 \over 2}}^2,
2\bar t_{-{1 \over 2}}t_{-{1 \over 2}},
t_{-{1 \over 2}}^2\}$
This warns us that to
establish any negative result such as no-decoupling, it is not
sufficient to establish it in the DS basis, but in {\it any}
permissible basis. In this connection it should be
observed that there is a new phenomenon that occurs because of the
 negative spin field, namely that there exist
{\it  scalar} combinations of the DS  fields. This means that
coefficients in any modification of the basis may be polynomials in these
scalars rather than constants. The scalars in question are
$\nu=\bar st^2_{-{3 \over 2}}$ for both models and
 $\mu=t_{-{3 \over 2}}\bar t_{-{1
\over 2}}$ for model $\hat \Gamma_q$.
\vskip 0.3truecm
 {\bf Deformation Structure of $W_d$.} We first
consider the subalgebras $W_d$ which have the spin content of
the canonical W-algebra. From the results in [4] there exists a
basis in which these algebra become
canonical for $t_{-{3 \over 2}}=0$ and $\bar v_{{1 \over
2}}=0$. So, at worst, they are deformations of $W_c$. The only question
is whether the deformation is trivial.
Fortunately there is a property of the present $G_2$ embedding
that permits us to determine this, namely that
the spin 1 part of the canonical W-algebra must be an $sl(2)$ KM algebra.
This means that a {\it necessary} condition for $W_d$ to be canonical is that
$W_1$ be an $sl(2)$ KM algebra. Hence to establish the contrary, it is
sufficient to
establish that there is no basis in which $W_1$
(which is much more tractable than $W_d$) forms an $sl(2)$ subalgebra.
\vskip 0.2truecm
 We first note that since the $\gamma_{{3 \over 2}}$
gauge-fixing determines the coefficient of $t_{-{3 \over
2}}v_{{1 \over 2}}$ in $s^c$ and $s^q$, we have
$$\{\bar s,s^c\}=\{\bar s,s^q\}=y+{1 \over 3}t_{-{3 \over 2}}^2\bar v_{{1 \over
2}}   \eqno(27)$$
which shows that there is no $sl(2)$ structure in the DS basis.
Thus we must consider the most general basis for the spin 1 sector
which is
$$\{\bar S,Y,S\} \qquad \hbox{where} \qquad \bar S=
\bar s +a\bar \tau^2\qquad Y=y+At\bar v_{{1 \over
2}}+\bigl(B\bar \tau +C\bar s t \bigr)u_{-{1 \over
2}}  \eqno(28)$$
and
$$\eqalign{
S=(s+{t \over 3} v_{{1\over 2}})+
Du^2_{-{1 \over 2}}+tEyu_{-{1 \over 2}}
&+t^2(F\Lambda +Gy^2 +Hu_{-{1 \over 2}}\bar v_{{1 \over
2}})  \cr
&+t^3(I\bar v_{{3 \over 2}}+Ky\bar v_{{1 \over 2}}) + t^4 N\bar
v^2_{{1 \over 2}}  \cr  } . \eqno(29)$$
where $t$ and $\bar \tau$ are abbreviations for
$t_{-{3 \over 2}}$ and $\bar t_{-{1 \over 2}}$ respectively, and
the coefficients $a,A,B$ etc. may be polynomials in the scalars
$\mu$ and $\nu$.
The expressions $v_{{1 \over 2}},\bar v_{{1 \over 2}}$ denote
$v_{{1 \over 2}}^c,\bar v_{{1 \over 2}}^c$ and
$u_{{1 \over 2}},\bar u_{{1 \over 2}}$ for the models
$\hat \Gamma_c$ and $\hat \Gamma_q$ respectively, and for the
model $\hat \Gamma_c$  we have
$D=-{1 \over 4}$ and $\mu=B=C=E=H=0$.

 It is important to note that in constructing (28) (29) we have restricted
ourselves to
polynomials which form a basis for the gauge-invariant polynomials.
i.e. to polynomials that remain unchanged in the linear limit.
Otherwise we would be considering not the freely-generated
polynomial W-algebra under consideration but the
algebras generated by various classes of functions
(general polynomials, rational functions etc.) of the DS
base-elements, which in general
have structures quite different from the W-algebra.
\vskip 0.2truecm
 The condition that the spin sector should form an $sl(2)$
KM algebra is that
$$\{\bar S,Y\}=\bar S \qquad \{\bar S,S\}=Y \qquad \{Y,S\}=S .
\eqno(30)$$
but since $\bar S$ is quadratic in $\tau$ and we shall finally be be
restricting ourselves to
configurations for which $\tau=0$,  we can simplify this condition by
letting $\bar S \rightarrow \bar s$. The $sl(2)$-condition then reduces to
$$\{\bar s,Y\}=\bar s \qquad \{\bar s,S\}=Y \qquad \{Y,S\}=S .
\eqno(31)$$
To investigate (31) we first compute some relevant
KM and W Poisson-brackets, namely
$$\{\bar s, t_m\}=\bar t_m  \quad \rightarrow \quad
\{\bar s,u_{-{1 \over2}}\}=\bar \tau \qquad
\{\bar s,v_{{1 \over2}}\}=\bar v_{{1 \over 2}} \qquad
\{\bar s,\bar v_{{3 \over2}}\}=-{1 \over 3}\bar s v_{{1 \over 2}}
\eqno(32)$$
 For the general basis the first condition in (31) is satisfied
automatically for the
$\hat\Gamma_c$ model and requires that $B=C=0$ for the model
$\hat \Gamma_q$. Since $\{\bar s,S\}$ must take the same general
form as $Y$ we see that the second condition simply gives linear
relations between the coefficients of $Y$ and $S$. The only relation
we shall need is the relation between the coefficients of $t\bar v_{{1
\over 2}}$ which is easily seen to be
$$A={1 \over 3}+Q \qquad \hbox{where} \qquad Q=H\mu+(K-{I \over
3})\nu-{1 \over 2}N\mu\nu  .\eqno(33)$$
The question then centres on the third condition in (31). Since $y$
alone reproduces $S$ the question is whether $\{\Delta y,S\}=0$ where
$\Delta y\equiv Y-y$. We shall show that this Poisson-bracket
is not zero by showing that it projects onto the Virasoro field
(or an equivalent gauge-invariant base-element).
For this it is sufficient to show that it projects onto $j_+$ since
such a base element is the only one in which $j_+$ can occur
linearly. As the only brackets that can produce $j_+$ are
$$\{t_{-{1 \over 2}},\bar t_{{3 \over 2}}\} \qquad
\{\bar t_{{1 \over 2}},t_{{1 \over 2}}\} \qquad
\{\bar t_{-{1 \over 2}},t_{{3 \over 2}}\}  \eqno(34)$$
and $\Delta y$ contains only $t_{-{1 \over 2}},\bar t_{-{1 \over
2}}$ and $\bar t_{{1 \over 2}}$ while $S$ does not contain
$t_{{3 \over 2}}$ we see that the coefficient of $j_+$ in
$\{\Delta y,S\}$ is
$$
{\partial \Delta y \over \partial \bar t_{{1 \over 2}}}
{\partial S \over \partial t_{{1 \over 2}}} +
{\partial \Delta y \over \partial t_{-{1 \over 2}} }
{\partial S \over \partial \bar t_{{3 \over 2}}}
={1 \over 3}(1-I\nu)At^2 +
\Bigl(B\mu +C\nu -{\epsilon \over 4}A\nu\Bigr)It^2
\eqno(35)$$
where $\epsilon=1,0$ for models $\hat \Gamma_c$ and $\hat
\Gamma_q$ respectively. Since this expression does not vanish even
for the configurations in which $\bar s=\bar \tau =0$ (for
which $\mu=\nu=0$ and hence $A={1 \over 3}$ from (33))  we
see that $\{\Delta y, S\}$ cannot be zero.
Thus there is no basis in which the spin 1 fields form an $sl(2)$
KM subalgebra and thus no basis in which $W_d$ is canonical.
\vskip 0.2truecm
 With hindsight, knowing that the projection of the
$W_1$-algebra onto the Virasoro occurs even for the configurations
$\bar s=\bar \tau=u_{-{1 \over 2}}=0$
and to order $t^2$ one can obtain the result in a faster and
more intuitive way. If we restrict to these configurations
we can omit in (28) and (29) all the terms that are quadratic
in these variables. If we restrict further to terms of order $t^2$
and recall that for constant fields the Virasoro term can be
neglected, we see that (28) (29) then reduce to
$$\bar S=\bar s \qquad Y=y+At\bar v_{{1 \over 2}} \qquad
S=s+{t \over 3}v_{{1 \over 2}}   \eqno(36)$$
which generalizes the DS basis only to the extent that it has
the free parameter $A$. The fact there is no $sl(2)$ structure
then follows from the observation
that there is no value of $A$ that satisfies both the second and
third conditions in (30).
\vskip 0.4truecm
{\bf No Decoupling} Although $W_d$ is only a deformation
of the canonical W-algebra it might still be possible for $W_d$
to decouple from its complement and some
further insight into the structure of the W-algebras of these models
is obtained by showing that this is not the case. A preliminary hint
that $t$ does not decouple from $W_d$ is that the Virasoro contains
a term $t_{-{3 \over 2}}\bar t_{{3 \over 2}}$ and thus there does not
seem to be any way in which the Virasoro could be split into a part
which acts only on $t$ and a part which acts only on $\bar v_{{3 \over 2}}$.
\vskip 0.2truecm
We first recall that the complement of $W_d$ consists of the
pairs of fields $\{t,\bar v_{{1 \over 2}}\}$ and $\{u_{-{1 \over 2}},\bar
\tau\}$ in the model $\Gamma_q$ and
the first of these pairs in the model $\Gamma_c$. These pairs of fields
are conjugate in the sense that
$$\{t,\bar v^c_{{1 \over 2}}\}={3 \over 2}(1+{\nu \over 4}) \qquad
\{t,\bar u_{{1 \over 2}}\}={3 \over 2}(1-\mu)\qquad \hbox{and} \qquad
\{\bar \tau,u_{-{1 \over 2}}\}=2(1-{\mu \over 4})   \eqno(37)$$
where the first two results refer to the models
models $\hat \Gamma_c$ and $\hat \Gamma_q$ respectively and the
third is relevant only for the model $\hat \Gamma_q$.
The fields in each conjugate pair are coupled by definition and in
fact they are intrinsically coupled in the sense that
no polynomial modification can make them decouple. This is because
any modification would add higher-degree terms and
the PB of $t$ with any such higher-degree terms would be at least
linear in the fields and thus could not compensate the constant
terms in (37). A new feature of the present models is that we then have
a negative grade field, namely $t$, intrinsically coupled to a
positive grade field. (This situation actually generalizes to any
DS model with  negative spin fields $W_{-s}$ because if such fields
survive the constraint and gauge-fixing conditions so do their
$M_-$ conjugate fields $W_{s+1}$, and the PB of the two conjugate elements
contains a non-zero constant).
\vskip 0.2truecm
Since the conjugate pairs are intrinsically coupled the question
then is whether $W_d$ can decouple from the conjugate pairs.
Let us first
consider the pair  $\{t,\bar v_{{1 \over 2}}\}$  which occurs in both
models. Using (37) and $\{t,u_m\}=-{3 \over 2}tu_{m-1}$ for $m\leq {1 \over
2}$ one finds that
$$\{t,Y\}=\Bigl((A(1+{\nu \over 4})-{1 \over 3}\Bigr){3t\over 2}
\quad \hbox{and} \quad
\{t,Y\}=\Bigl( (A-{1 \over 3})-(A+B)\mu -C\nu\Bigr){3t \over 2}
\eqno(38)$$
for the models $\hat \Gamma_c$ and $\hat \Gamma_q$ respectively.
{}From (38) we see that in the first model there is no polynomial value
of the parameters for which $t$ decouples.
\vskip 0.2truecm
In the second model $t$ decouples from $Y$ for
$A=-B={1 \over 3}$ and $C=0$ i.e. for
$Y=y+{t \over 3}\bar u_{{1 \over 2}} -{\bar \tau \over 3}u_{-{1
\over 2}}$ and one can check that it also decouples from
$S=s+{t \over 3}u_{{1 \over 2}}-{1 \over 6}u^2_{-{1 \over 2}}$.
So there is a choice of parameters for which $t$ does not
couple directly to the spin 1 sector. However, for the conjugate
field, which reduces to $\bar u_{{1 \over 2}}$ in this model, we have
$$\{Y,\bar u_{{1 \over 2}}\}=
\{y+{t \over 3}\bar u_{{1 \over 2}}-{\bar \tau \over 3}u_{-{1 \over
2}},\bar u_{{1 \over 2}}\}=
-{\mu  \over 2}\bar u_{{1 \over 2}}+...\eqno(39)$$
where the dots denote terms not containing $\bar u_{{1 \over 2}}$,
which shows that $\bar u_{{1 \over 2}}$ couples to $Y$. Furthermore,
this coupling cannot be compensated by a polynomial modification
of $\bar u_{{1 \over 2}}$ because the most general polynomial
modification is
to add terms of the form $y\bar \tau$, $\bar su_{-{1 \over 2}}$ and
$\bar syt$ and the part of the PB's of these terms with $Y$ which
contain $\bar u_{{1\over 2}}$ vanish on the surface
$\bar s=y=0 $ where the coefficient of $\bar u_{{1
\over 2}}$ in (39) does not vanish. Thus in the model $\hat \Gamma_q$
the field $\bar u_{{1 \over 2}}$ couples directly to the spin 1
sector and the field $t$ then couples indirectly.
\vskip 0.2truecm
Let us next consider the  conjugate
pair $\{\bar \tau,u_{-{1 \over 2}},\}$, which occurs only
for the model $\hat \Gamma_q$. This pair is of interest because it decouples
in the quasi-canonical model $\Gamma_q$. To see whether it decouples in the
model $\hat \Gamma_q$  we compute the PB of $\bar s+a\bar \tau^2$
and $u_{-{1 \over 2}}+byt$, which are the most general fields
that can be constructed corresponding to the base elements
 $\bar s$ and $u_{-{1 \over 2}}$. We obtain
$$\{\bar s+a\bar \tau^2,  u_{-{1 \over 2}}+byt \}=
[(1+4a)\bar +a(b-1)\mu]\bar \tau  +b\bar st  . \eqno(40)$$
In the quasi-canonical case for which $t=0$ this vanishes for
$4a=-1$ but in the present case there is clearly no polynomial
choice of parameters for which it vanishes. Thus there is no basis
in which this pair decouples from $W_d$.
\vskip 0.2truecm
 From the above results we conclude that for neither of
the present models is there a basis in which $W_d$ decouples
from its complement.
\vskip 0.2truecm
{\bf Summary and Comments:} We have constructed two
W-algebras by the reduction of a $G_2$ Kac-Moody algebra, using
a non-principal $sl(2)$ embedding and constraint algebras which
are $sl(2)$
positive and are proper subalgebras of the canonical constraint
algebra. Both W-algebras have a field of strictly negative conformal
weight and a subalgebra $W_d$ which is a
deformation of the canonical W-algebra. The subalgebra $W_d$ does
not decouple from its complement or even from the negative spin
field. Indeed the W-algebra
would appear to be indecomposable. We have not given the details of
the W-algebra, partly because this could be done easily only in the
DS basis which is not primary and is not
expected to exhibit the structure in its simplest form and partly
because the details are not of any great interest at present.
The primary basis has to be constructed from the DS basis in an
ad hoc manner and an interesting question is to whether one could
find a systematic procedure for obtaining the primary
basis directly for this kind of model. The appearance of negative spin
fields which do not decouple is a little surprising as it would seem to imply
that the quantized version of the theory could not be unitary (the
Virasoro operator could not be self-adjoint [5]) but it remains to be
seen whether this is really the case. Finally it should be mentioned that
the results would be essentially the same even if the Virasoro
operator were not unique since the reduced algebras would still be
W-algebras with respect to $\Lambda$ and the non-decoupling
and $sl(2)$ non-closure results are statements about base elements
which do not depend strongly on the spin-labelling.
\vskip 0.3truecm
{\it Acknowledgement} \hskip 0.2truecm The authors wish to thank Dr. I.
Tsutsui and the referee for some valuable criticisms and suggestions.
\vskip 0.4truecm
\noindent {\it References}
\vskip 0.2truecm
\noindent [1] L. Feher, L. O'Raifeartaigh, P. Ruelle, I. Tsutsui and A.
 Wipf, Physics Reports {\bf 222}(1992)1-64  and references therein.

\noindent [2] P. Bouwknecht and K. Schoutens, Physics Reports {\bf
223}(1993)183  and references therein.

\noindent [3] L. Feher, L. O'Raifeartaigh, P. Ruelle and I. Tsutsui,
Comm. Math. Phys. (in press)

\noindent [4] J. de Boer and T. Tjin, Comm. Math. Phys. {\bf
160}(1994)317 P. Bowcock and G. Watts, Nucl. Phys.
{\bf 379B}(1992) 63-96

\noindent [5] P. Goddard and D. Olive, Int. J. Mod. Phys. {\bf A1}
(1986) 303.

\bye